\documentclass{svproc}
\usepackage{url}
\usepackage{xurl}
\usepackage{array}
\usepackage{hyperref}
\usepackage{breakurl}
\usepackage{graphicx}
\usepackage{float}

\begin{document}
\mainmatter              
\title{SISA: Securing Images by Selective Alteration}
\titlerunning{SISA: Securing Images by Selective Alteration}  
%
\author{Prutha Gaherwar\inst{1} \and Shraddha Joshi\inst{1} \and
Raviraj Joshi\inst{2} \and Rahul Khengare\inst{1} }
\authorrunning{Prutha Gaherwar et al.} 
%
\tocauthor{Shraddha Joshi, Raviraj Joshi, Rahul Khengare}
\institute{Department of Computer Technology, Pune Institute of
Computer Technology, Pune, Maharashtra, India\\
\email{pruthagaherwar@gmail.com}\\
\email{joshishraddha597@gmail.com}\\
\email{rahulk1306@gmail.com}\\
\and
Department of Computer Science and Engineering, Indian Institute of
Technology Madras, Chennai, Tamil Nadu, India\\
\email{ravirajoshi@gmail.com}}

\maketitle              

\begin{abstract}
With an increase in mobile and camera devices' popularity, digital content in the form of images has increased drastically. As personal life is being continuously documented in pictures, the risk of losing it to eavesdroppers is a matter of grave concern. Secondary storage is the most preferred medium for the storage of personal and other images. Our work is concerned with the security of such images. While encryption is the best way to ensure image security, full encryption and decryption is a computationally-intensive process. Moreover, as cameras are getting better every day, image quality, and thus, the pixel density has increased considerably. The increased pixel density makes encryption and decryption more expensive. We, therefore, delve into selective encryption and selective blurring based on the region of interest. Instead of encrypting or blurring the entire photograph, we only encode selected regions of the image. We present a comparative analysis of the partial and full encryption of the photos. This kind of encoding will help us lower the encryption overhead without compromising security. The applications utilizing this technique will become more usable due to the reduction in the decryption time. Additionally, blurred images being more readable than encrypted ones, allowed us to define the level of security. We leverage the machine learning algorithms like Mask-RCNN (Region-based convolutional neural network) and YOLO (You Only Look Once) to select the region of interest. These algorithms have set new benchmarks for object recognition. We develop an end to end system to demonstrate our idea of selective encryption.
\end{abstract}
\section{Introduction}
\par Data privacy is a crucial part of the digital era. Businesses and individuals must give a significant amount of thought to how data is gathered, retained, applied, and disclosed. When there is unauthorized access to private or personal information, a privacy breach occurs. It often leads to legal action if the violation is severe. Every business unit, government, and private sector frequently uses the digital image to transfer critical data. These images accessible over the internet may not be secure; therefore, image security has always been an important subject. A widely used solution to the data breach is encryption, which encodes data with encryption algorithms so that even unauthorized access cannot derive meaningful information\cite{mohammad2017survey}.
\par While encryption is the best technique to safeguard our images, full encryption may not always be required\cite{khatod2020enigma}. For example, a family photograph may contain background scenery, which may not be necessary. It would be sufficient to encrypt only the people in the images. Therefore, encryption hides the unnecessary data along with the necessary one, which increases the processing time. We propose SISA, which explores the idea of partial encryption\cite{yu2016iprivacy} to reduce this overhead. Our work is concerned with the security of images which are mostly stored in the secondary storage. Although we mainly discuss pictures stored on our hard drives, it is equally applicable to cloud storage.\\
\begin{figure}[h]
    \centering
    \includegraphics[width=0.9\textwidth]{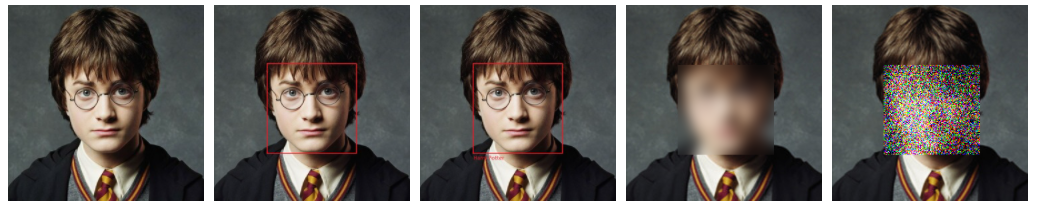}
    \caption{SISA results on an image. The left-most image is the original image; the next image is annotated with a bounding box. The image at the center is the output of the face recognition algorithm used (you can see the person's name on the bottom-left of the bounding box). Next image is an output of the blurring technique (we have elaborated the selective blurring using instance segmentation in the later sections), and the right-most image is the selectively encrypted image.
}
    \label{fig:hp1}
\end{figure}

\subsection{Importance of Security}
Every year, even the common man generates exabytes of data just in the form of images. These images comprise personal data like copies of passport, social security documents, bank details, or biodata. So the prospect of image theft resulting in the loss of important personal data is an extremely serious problem at hand.
\par The security measures to safeguard critical information are implemented at multiple levels. On one side, we have machine-level security employing firewalls. Whereas on another, we have storage level security that uses authentication and authorization as safety measures. However, it is very much possible to break into machine-level security and single-step storage security. In a survey by Varonis\cite{110MustK90:online}, security breaches have increased by 11\% since 2018 and 67\% since 2014. It is too big of a percentage to be ignored or side-lined. 
\par In notable cases, information leakage completely invades personal privacy. For example, the malicious spread of photos in private online albums or patients' medical diagnosis images may cause numerous losses to the individual or society. According to IBM\cite{IBMStudy19:online}, healthcare spent the most time in the data breach lifecycle in 2019, 329 days. As the essential data transfers are carried out extensively using digital photographs, safety is the primary concern for a system to maintain integrity, confidentiality, and authenticity.

\subsection{Proposal}
The widely used technique of complete image encryption while storing the data has some significant disadvantages. These techniques encrypt each pixel one by one, so when the pixel count rises, the encryption time increases as well \cite{mohammad2017survey}. Besides, the image quality has evolved. If we consider an HD image, which is the most preferred quality these days, it has 20L pixels (1920 x 1080), whereas a 4k image has whooping 80L (3840 x 2160) pixels. Encrypting and decrypting 20-80L pixels will be a too time-consuming process. Moreover, the encryption and decryption algorithms are quite resource-intensive, making it difficult to run on low-end devices. So an application encrypting and decrypting bulk images on a local device will be far from usable.
\par This paper presents an approach based on partial blurring and encryption. The aim is to ensure image security while reducing the overhead of processing time. SISA detects the objects in the image and assigns them a priority value based on user preference to choose the most critical area. The default priority value is set as per the affinity towards the center coordinates of the image, i.e., in general cases, the highest priority object will be located at the center of the image, and the one with the least priority will be at the edges. Next, we apply selective encryption to the file (e.g., jpg, png) which encrypts only parts of data instead of encrypting the entire data. In essence, with our approach we try to combine the benefits of prioritization and encryption.\\\\
The main contributions in our paper are as follows:
\begin{itemize}
  \item Our scheme identifies the significant parts of the user data based on the prioritization technique, which we later use for selective alterations of the image.
  \item We focus on selective alteration such that in event of a successful attack, the meaningful information is not visible to the attacker. Also, we compare full and partial encryption and highlight the challenges from a deployment perspective.
\end{itemize}

\subsubsection{Architecture.}

\par The image to be secured is passed to the SISA layer before storing it on disk. The layer appropriately modifies the image based on preferences selected by the user. One of the strong user preferences is the level of security required. The image is encrypted for higher security levels, whereas, for lower levels, the image is blurred. Although blurring is used with low levels of security, it is imperative from a usability perspective. A blurred image is more visually accessible as compared to an encrypted image and hence more usable.
\begin{figure}[h]
    \centering
    \includegraphics[width=0.75\textwidth]{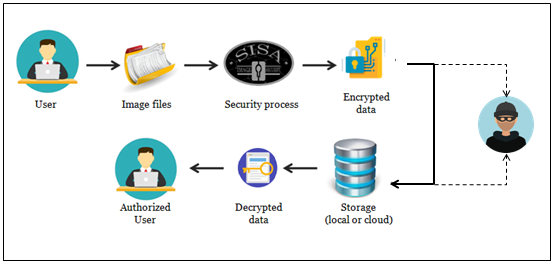}
    \caption{High-level architecture}
    \label{fig:arch}
\end{figure}
\par When these images are transferred through an insecure channel, even if the intruder eavesdrops on the data on the wire, no usable information will be revealed. These images will also be present in the storage in the modified format, thus hacking into the storage will also render the data unusable. Only the authorized user will be able to access the valid data using a security key.
\begin{figure}[h]
    \centering
    \includegraphics[width=0.75\textwidth]{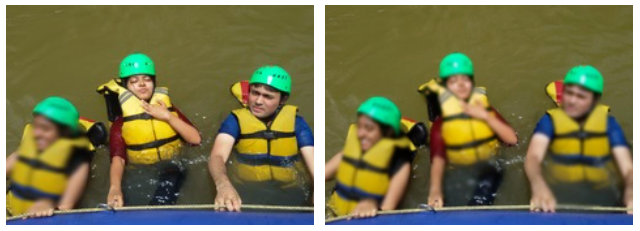}
    \caption{The percentage of blurring in the left image is 30, and in the right image, it is 70. Here we can see that the object of highest priority (user's image) is blurred in the left image, while in the right, other people are also blurred.
}
    \label{fig:lvlblur}
\end{figure}
\par The level of security also controls the percentage of the image to be encrypted. For the lowest level, 30\% of the image is encrypted. This percentage is linearly increased for higher levels of security. The benefit of our approach is the intelligent selection of 30\% region. (refer Figure \ref{fig:lvlblur}). Rather than randomly selecting pixels, we use region selection and prioritization algorithms to select relevant pixels.

\subsubsection{Process flow.}
\begin{figure}[h]
    \centering
    \includegraphics[width=0.75\textwidth]{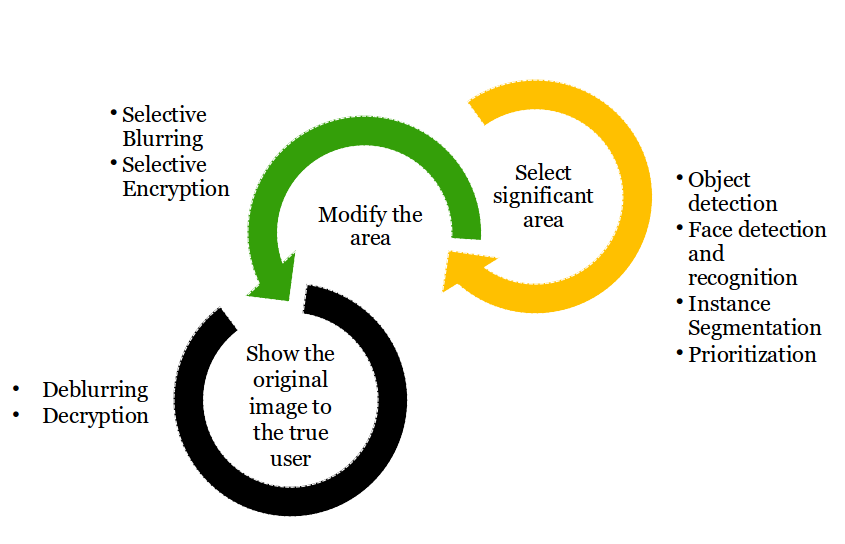}
    \caption{Process flow}
    \label{fig:processflow}
\end{figure}
As shown in the Figure \ref{fig:processflow}, the process flow will be as follows:
\begin{enumerate}
     \item Select a significant area
     \begin{enumerate}
       \item As the first step, the algorithm of object detection is used to produce bounding-boxes of objects in the picture.
       \item The Mask R-CNN algorithm outlines the object's edges with a bounding box and marks the object instance.
       \item The prioritization scheme then assigns a priority to these objects. The priority is based on user preferences. Simple heuristics like importance to the foreground, background, or face are part of the preferences.
       \item In addition to this, we also carry out text detection, face detection, and face identification to select a significant area. These extra selections are part of user preferences and appropriately triggered.
     \end{enumerate}
     \item Modify the area under consideration
     \begin{enumerate}
       \item For a low level of security, the image is blurred, and for a higher level, it is encrypted. The level of security also controls the percentage of the image regions to be altered.
       \item As per the selections, all these modifications are carried out by altering the image's required part.
     \end{enumerate}
     \item Display the original image
     \begin{enumerate}
       \item The original image can be retrieved using a de-blurring or decryption algorithm.
     \end{enumerate}
\end{enumerate}

\section{Methodology}
\subsection{Processing on the image}
\subsubsection{Object detection.}
It is a classification and localization problem. Generally, we can not fix on the number of objects in the image. The model expects data normally in the form 
(I, A), where I is an image, and A is an array containing (class\_label, p, q, l, b).\\\\For A = (class\_label, p, q, l, b);\\
p \hspace{1mm}=\hspace{1mm} x-coordinate of top left corner of the bounding-box\\
q \hspace{1mm}=\hspace{1mm} y-coordinate of top left corner of the bounding-box\\
l \hspace{1.9mm}=\hspace{1mm} bounding-box length in pixel\\
b \hspace{1mm}=\hspace{1mm} bounding-box breadth in pixel\\\\
The necessary steps followed are:
\begin{enumerate}
\item Take an input image
\item Extract region maps
\item Compute attributes of each submission using an extensive convolutional neural network
\item Classify each region using class-specific linear SVM (Support Vector Machine)
\end{enumerate}

\subsubsection{Face detection and recognition.}
The face recognition consists of a two-step process:\\
(1) face detection (bounded face) in image\\
(2) face identification (person identification) on the detected fixed face\\\\
The face detection algorithm detects all the faces in the image and then puts a bounding box around it. Face identification generates the detected bounded face's embedding and matches it against the embedding of the training faces in the user-specific database.

\subsubsection{Instance segmentation.}
Instance-segmentation aims to find specific objects in an image and develop a mask around the object. Instance-segmentation can be considered as object detection where the output is defined as a object mask. This is in contrast to normal object detection where the output is a bounding box. Unlike semantic segmentation, which tries to classify each pixel in the image, instance-segmentation does not label every pixel in the picture\cite{jiao2019survey}.
\par Several enhancements over R-CNN for object detection has resulted in Mask R-CNN\cite{he2017mask}. R-CNN applies two step process to generate an object-label and the bounding-box. Firstly, it adopts selective search method to create region maps and then each obtained region is transformed. Fast R-CNN\cite{girshick2015fast} is an improved R-CNN with a further improved version labeled as Faster R-CNN\cite{ren2015faster}. 
The Mask R-CNN algorithm is an enhancement over Faster R-CNN by parallelizing a mask predicting branch with the class-label and bounding-box prediction branch. Despite of increasing a tiny expense over the Faster R-CNN network, it still provides optimum speed and is the most efficient of all.

\subsubsection{Prioritization.}
The prioritization step involves assigning priorities to the objects detected by the object detection and instance segmentation algorithms. In day to day life, while taking a picture, we usually keep objects of most importance in the center of the frame. Using this generic idea, we developed an algorithm that selects objects closest to the image's center.
The interactive learning of the critical regions can improve this default scheme according to the user specifications.
\begin{figure}[h]
    \centering
    \includegraphics[width=1\textwidth]{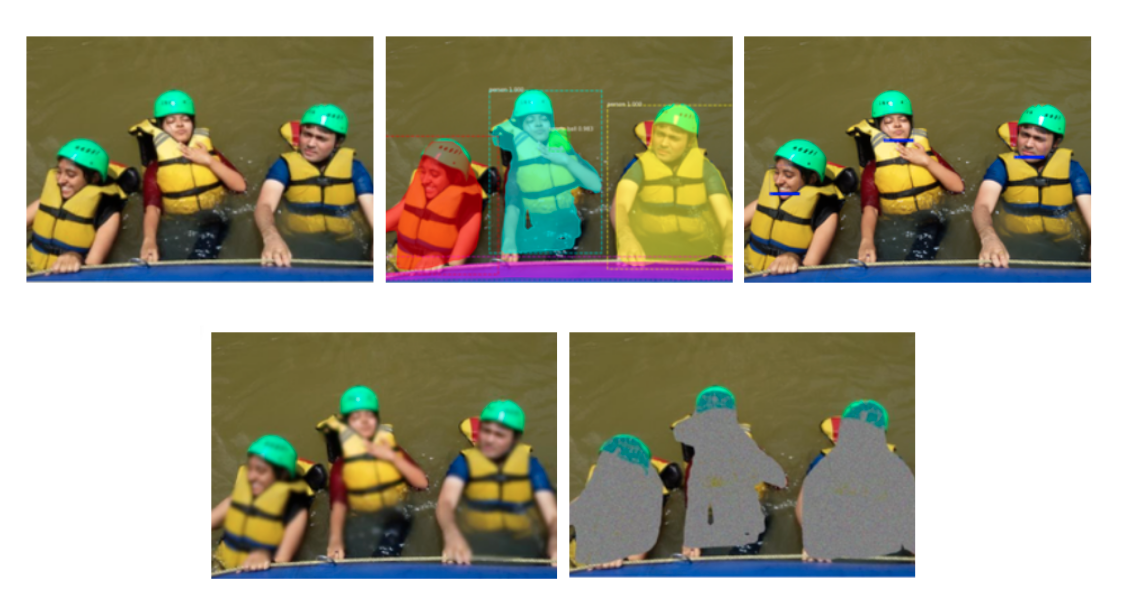}
    \caption{The first image is the input on which we have generated the following outputs. The second image shows the bounding boxes and segmented instances. The faces of the people that can be seen in the picture are detected and recognized in the next image. The bottom-left image is the output of the blurring algorithm, and the last image is the output of the encryption algorithm.}
    \label{fig:result}
\end{figure}

\subsection{Modifying the image}
\subsubsection{Blurring and deblurring.}
Gaussian blur is carried out using a convolution kernel. We prepare a 2D matrix of normalized pre-calculated Gaussian function values. This matrix is used as weights to calculate weighted sum of all adjacent pixels to find new value of each pixel\cite{filip2014investigation}.\\ Partial blurring is achieved by applying Gaussian blur on the area covered by prioritized objects.
\begin{eqnarray*}
  G(x,y)= \frac{1}{2\pi\sigma^2}(e^-\frac{x^2+y^2}{2\sigma^2})
\end{eqnarray*}

\subsubsection{Encryption and decryption.}
Encryption is the process of encoding a simple image into a ciphered image using encryption algorithms so that unauthorized users cannot access the original image. Furthermore, the process of recovering an image from a ciphered image using a set of instructions is called decryption. It is carried out using a secret key. \\
We use the standard AES algorithm with CFB mode for encryption. The selected group of pixels are modified to form an encrypted image. The standard process is used for decryption as well.

\subsubsection{Metadata.}
The information required to reconstruct the image is stored in the image's metadata using Exif-tool. Trivially this includes the encrypted portion's coordinates. Any mismatch in encrypted contents length is also added to the metadata.\\
In Figure \ref{fig:result}, we have shown the implementation of all the techniques used.
\section{Literature Survey}
\begin{enumerate}
\item DeepPrivacy: A Generative Adversarial Network for face Anonymization:
The paper introduces an innovative model that automatically anonymizes faces present in the pictures while preserving the primary data distribution. It ensures the complete anonymization of each face seen in the picture by creating the images solely based on the privacy-safe data. Their model is established on a conditional generative adversarial network, developing photos by taking the existent pose and image environment into consideration.\cite{hukkelaas2019deepprivacy}\\

\item You Only Look Once (YOLO):
YOLO presents the unified model for the detection of objects. It spatially separates objects' bounding-boxes and related class probabilities, viewing object detection as a regression problem. As the detection pipeline comprises only a single network, end-to-end optimization based on detection performance is possible. The convolutional layers are pre-trained using the ImageNet 1000-class competition dataset. An accuracy of 88\% was achieved.\cite{Redmon_2016_CVPR}\\

\item Fast Face-swap using convolutional neural networks:
This paper presents a feed-forward neural network that achieves high realism levels in generated face-swapped images. This method uses a multi-image style loss, thus approximating a manifold describing a style rather than using a single reference point. The trained networks allow the model to perform face swapping in real-time.\cite{korshunova2017fast}\\

\item Techniques for selective encryption of uncompressed and compressed images:
This paper proposes two approaches for selective encryption of the image. The first approach focuses on raster images. The second approach adapts the JPEG compression scheme, which puts forth, one,  a constant bit rate and another, format compliance. It involves encryption of the sign and magnitude of non-zero DCT coefficients. This approach also elaborates on its usage for several selective encryptions.\cite{van2002techniques}\\\\
\end{enumerate}

\begin{table}[H]
\caption{Findings from the Literature Survey} 
\begin{center}
\begin{tabular}{ | m{11em} | m{11em}| m{11em} | }
\hline
Title & Pros & Cons \\
\hline
DeepPrivacy: A Generative Adversarial Network for face Anonymization\cite{hukkelaas2019deepprivacy} &
\begin{itemize}
    \item [$\bullet$]Automatic anonymization of all faces.
    \item [$\bullet$]Loss of mask allows class prediction without competition.
\end{itemize}
 &
 \begin{itemize}
     \item [$\bullet$]Can generate unrealistic images
 \end{itemize}
\\
\hline
You Only Look Once (YOLO)\cite{Redmon_2016_CVPR} 
&
\begin{itemize}
\item [$\bullet$]A single network completes the whole detection pipeline.
\end{itemize}
&

\begin{itemize}
\item [$\bullet$]Can miss small objects which appear together.
\item [$\bullet$]Can fail for unidentified objects, aspect ratios , or configurations
\end{itemize}
\\
\hline
Fast Face-swap Using Convolutional Neural Networks\cite{korshunova2017fast}
&
\begin{itemize}
\item [$\bullet$] Feed forward NN achieves photorealism
\item [$\bullet$]Style transfer
\end{itemize}

&
\begin{itemize}
\item [$\bullet$]Quality depends on the collection of styled images.
\end{itemize}
\\
\hline
Techniques for selective encryption of uncompressed and compressed images\cite{van2002techniques}
&
\begin{itemize}
\item [$\bullet$]Constant bit rate and format compliance
\end{itemize}
&
\begin{itemize}
\item [$\bullet$]JPEG specific
\item [$\bullet$]Applicable only to monochrome images
\end{itemize}

\\
\hline
\end{tabular}
\end{center}
\end{table}

\section{Mathematical Model}
\begin{eqnarray*}
  S = \{s, e, X, Y, DD, ND, ff, fme\}
\end{eqnarray*}
$I = \{i | i$ is the image uploaded by user\}\\
$P = \{ p | p$ is tuple of object properties in an image\}\\
$Tuple =$ \{label, coordinates of the top-left corner, width, height, mask\}\\
$Z = \{z | z$ is the tuple of label and priority\}\\
$O = \{o | o$ is the altered image\}\\\\
$s:$ Initialize
\par $I=\Phi, P=\Phi, Z=\Phi, O=\Phi$\\\\
$e:$ Update prioritization model\\\\
$X:$ Input
\par $I, l, k \in X$
\par $l:$ level of security
\par $k:$ security key\\\\
$Y:$ Output\\\\
$DD:$ Deterministic data\\\\
$ND:$ Non-deterministic data
\par $ND = \Phi$
\par $O \in Y$\\\\
$ff:$ Friend function
\par $f1(x) :$ Object detection function
\par $I \rightarrow P$
\par $f2(x) :$ Alteration function
\par $Z, P, k, l \rightarrow O$
\par As per the value of l and the sets present, blurring/encryption/morphing \par takes place\\\\
$fme:$ Generated Function
\par $f3(x) :$ Prioritization function
\par $P \rightarrow Z$

\section{Results}
\par The COCO dataset is used to train the Mask R-CNN model, and custom made data set to train the face recognition algorithm. For example, to generate the results of Figure \ref{fig:hp1}, we trained the face recognition model on the subject’s photos taken from various angles. The performance estimation of existing encryption and SISA algorithms is done on a machine with a 64bit OS and an 8GB RAM.\\
\begin{figure}[h]
    \centering
    \includegraphics[width=0.9\textwidth]{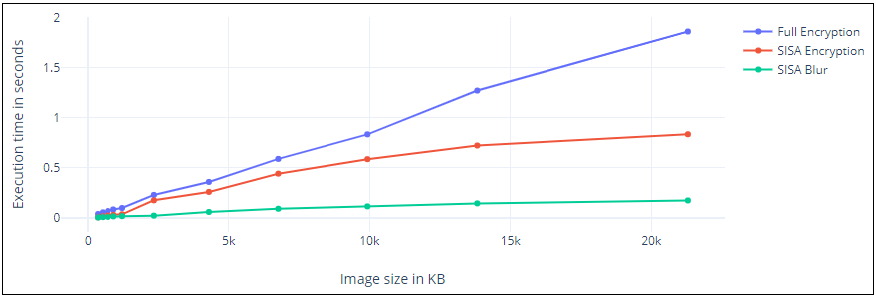}
    \caption{Full Encryption Vs. SISA Encryption Vs. SISA Blur}
    \label{fig:grf1}
\end{figure}
\begin{figure}[H]
    \centering
    \includegraphics[width=0.9\textwidth]{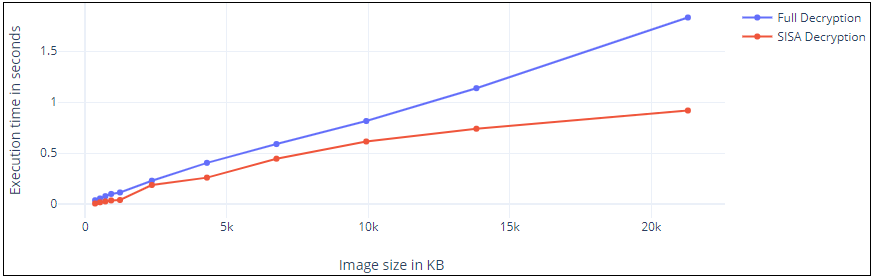}
    \caption{Full Decryption Vs. SISA Decryption}
    \label{fig:grf2}
\end{figure}

\par From a deployment perspective, decryption is more critical than encryption. The reconstruction time of the image should be less as it will determine the usability of the application. Conversely, encryption, and backup can be done in the background without exposing this latency to the user. To show the effectiveness of selective encryption in terms of usability, we compare our approach with full encryption and decryption. Figure \ref{fig:grf1} shows the comparison of encryption or blurring times for different approaches. Figure \ref{fig:grf2} shows the corresponding timing required during decryption. The time needed for blurring is significantly less as compared to encryption and decryption. So if we can compromise on the level of security, then blurring is a good option. However, blurring is a lossy operation, and hence the blurred object needs to be stored separately to reconstruct the original image precisely. The objects stored separately are also encrypted, so deblurring time is equivalent to decryption time. We show that as the size of the image increases, the processing time for altering the complete image increases substantially; comparatively, the time required for SISA is almost 46\% less.

\section{Conclusion}
In this paper, we present an approach for securing images by selective alteration. Here, we encode an image by selective encryption or blurring, which cuts down the time overhead without compromising security. As the image is encrypted with various constraints and modification types for each image, it becomes difficult for the intruder to access the original image. Although blurring is a less secure alternative, it is comparatively more user friendly. When integrated with security applications, SISA can provide an efficient interface for secure data storage. Due to partial encryption, the data's actual image recovery time will be substantially less, and these applications will be more serviceable.

\section{Application and Future Scope}
SISA can be used to secure banking documents, digital signatures, personal confidential snapshots, and medical records. It can also be used with personal identification documents like PAN card and Aadhar card. Moreover, it can be incorporated in both cloud-based or local image sharing applications. With the popularity of mobile cameras, photographs are a very convenient way to store capture information these days. So, any critical image with confidential content can be passed through the SISA layer before storing it on disk or cloud. SISA can be further enhanced by including morphing of images and text encryption inside the image. In the context of medical records like X-rays and MRI scans, we will need a more specific object detector as compared to our current general object detection approach. Currently, we rely on user preferences to rank the objects for encryption or blurring. It can be fully automated to have an end to end machine learning-based pipeline. 

\section*{Acknowledgements}
We would like to take this opportunity to show our gratitude towards the L3Cube group, which sponsored the project and guided us throughout, and credit one of our teammates, Garvit Gulati, who was also a part of this project's implementation.

\bibliographystyle{styles/bibtex/splncs03}
\bibliography{main}
\end{document}